\documentclass[twocolumn,nofootinbib]{nature}

\usepackage{amsmath}    
\usepackage{color}
\usepackage[10pt]{moresize}

\usepackage{graphicx}
\usepackage{caption}
\DeclareCaptionLabelFormat{adja-page}{\hrulefill\\#1 #2 \emph{(previous page)}}
\usepackage{subfig}

\usepackage{amsfonts}
\usepackage{amssymb}
\usepackage{amscd}
\usepackage{enumerate}
\usepackage{epsfig}
\usepackage{epstopdf}
\usepackage{dcolumn}
\usepackage{bm}
\usepackage{amsthm}
\usepackage{amsfonts}
\usepackage{color} 
\usepackage[usenames,dvipsnames]{xcolor}
\usepackage{amsmath}
\usepackage{enumerate}
\usepackage{bbm}
\usepackage[colorlinks=true,citecolor=blue,urlcolor=black]{hyperref}

\begin{document}

\title{Practical decoy-state round-robin differential-phase-shift quantum key distribution}

\author{Ying-Ying Zhang$^{1,2}$, Wan-Su Bao$^{1,2,}$$^\star$, Chun Zhou$^{1,2}$, Hong-Wei Li$^{1,2}$,  Yang Wang$^{1,2}$,  Mu-Sheng Jiang$^{1,2}$}

\maketitle

\begin{affiliations}
\item Zhengzhou Information Science and Technology Institute, Zhengzhou, 450001, China
\item Synergetic Innovation Center of Quantum Information $\&$ Quantum Physics, University of Science and Technology of China, Hefei, 230026,
China\\
$^\star$e-mail:2010thzz@sina.com.
\end{affiliations}

 \baselineskip24pt

\maketitle

\begin{abstract}

To overcome the signal disturbance from the transmission process, recently, a new type of protocol named round-robin differential-phase-shift(RRDPS) quantum key distribution[Nature 509, 475(2014)] is proposed. It can estimate how much information has leaked to eavesdropper without monitoring bit error rates. In this paper, we compare the performance of RRDPS using different sources without and with decoy-state method, such as weak coherent pulses(WCPs) and heralded single photon source(HSPS). For practical implementations, we propose finite decoy-state method for RRDPS, the performance of which is close to the infinite one. Taking WCPs as an example, the three-intensity decoy-state protocol can distribute secret keys over a distance of 128 km when the length of pulses packet is 32, which confirms the great practical interest of our method.

\end{abstract}

\section*{Introduction}

Quantum key distribution(QKD)\cite{1, 2} enables two legitimate communication participants, Alice and Bob, to share identical keys based on the fundamental laws of quantum physics. It has been a kind of information security technology raised from modern cryptography and quantum mechanics. QKD can be considered the most important application of quantum physics. Many papers have proven the security of Bennett-Brassard-1984(BB84) protocol\cite{3, 4}. Subsequently the experimental implementation also made great progress\cite{6, 7, 9}. So far, commercial products have already become available. However, due to the symmetry of BB84 protocol, the phase error rate is equal to the bit error rate. Needing to estimate the amount of leaked information by randomly sampling the signal, BB84 protocol may overestimate the leaked information and limit the threshold of the error rate.

Recently, a quantum key distribution protocol called round-robin differential phase-shift(RRDPS) was proposed by Sasaki et al.\cite{11}, which can generate keys without monitoring signal disturbance of the measurement outcomes and has no restriction on the error rate\cite{12}. RRDPS is a novel method to encode raw key bits even with the existence of eavesdropper, in which Bob specifies randomly how to calculate the sifted key from the raw key bits. Due to the large number of pulses in a packet, RRDPS system has higher stability and lower loss. It can tolerate a noisier channel than the conventional one. Up to now, many modified schemes have been proposed and several experimental demonstrations have been performed\cite{13, 14, 15, 16}. Using a receiver set-up to randomly choose one of four interferometers with different delays, Takesue et al.\cite{13} reported a proof-of-principle QKD experiment based on RRDPS protocol. Wang et al.\cite{14} demonstrated  an active implementation of this protocol, and their system can distribute secret keys over a distance of 90 km. Implementation results show that the protocol is feasible with current technology, especially in high-error situations\cite{14}.

Similar to the standard QKD protocol, most researches of RRDPS use weak coherent pulses(WCPs) as a replacement of the perfect single-photon source. Heralded single photon source (HSPS), is also within reach of current technology and can be considered as the candidate of the perfect single-photon source. So the performance of this source in RRDPS remains to be further studied.

In BB84 protocol, decoy-state method\cite{17, 18, 19, 20} can be used to efficiently estimate the contribution of the single-photon pulse and significantly increase the transmission distance of secure keys. Many researchers have studied the practicability of decoy-state method. Ma et al. \cite{20} first studied the statistical fluctuation analysis for the decoy state. Similarly, Zhang et al.\cite{12} proposed the infinite decoy-state method for RRDPS. Their method is valid in the asymptotic limit with an infinite number of decoy states, thus it has some limitations in practice. We extend it to the practical case with a finite number of decoy states.

First of all, we compare the performance of WCPs and HSPS in different pulse packet lengths. Then, fixing the packet length at 32, we simulate the infinite decoy-state method using these two sources. Since infinite decoy-state method is difficult to achieve in practice, we put forward a finite decoy-state protocol. Taking WCPs as an example, bounds on the yields and quantum bit error rates of some photon number states are stated. At last, considering that contributions from three and more photons are not obvious, we employ three-intensity decoy-state method.

\section*{Results}
\textbf{Round-robin differential-phase-shift protocol}

The round-robin differential-phase-shift protocol encodes raw key bits coherently so that only a few bits can be read out at the same time. It is hard for Eve to guess the sifted key. The ideal protocol, between two legitimate users, Alice and Bob, runs as follows.

1. State Preparation. Alice prepares packets of pulses containing $L$ pulses, and generates a random  $L$-bit sequence, $s=(s_1, s_2,\cdots, s_L)$. Then she encodes the sequence into the phase of each pulse, 0(when $s_i=0$) or $\pi$(when $s_i=1$), and sends the pulse packets to Bob. We consider that the encoded signal is a superposition photon state of  pulses\cite{11}
\begin{equation}
|\psi>=\frac{1}{{\sqrt L}}\sum\limits_{k = 1}^L {{{(-1)}^{{s_k}}}}|k>
\end{equation}
where ${s_k}$ is the encoded bit sequence, $|k>$ denotes that the photon is in the $k$-th pulse.

2. Measurement. Upon receiving the states, Bob randomly sets the pulse delay value $r(1\le r\le L - 1)$, and splits each received L-pulse train into two. Then Bob delays one of the train and interferes with the other. After measuring interference between the two trains, the detection result shows the phase difference between two pulses  $i$ and$j$ $(\{i, j\}\subset\{1, 2, \cdots L\})$  satisfying $j-i=\pm r(\bmod L)$. The value of the relative phase ${s_B}={s_i}\oplus {s_j}$  is a sifted key of Bob.

3. Sifting. Bob announces  $\{i, j\}$ to Alice through classical channel, so that Alice computes ${s_A} = {s_i} \oplus {s_j}$  as her sifted key.

4. Post Processing. Alice and Bob repeat steps 1-3 to accumulate enough sifted key. They perform error correction and privacy amplification on the sifted key to extract the final secure key.

The sketch of the protocol is shown in the Figure 1.

\begin{figure}[ht]
\centering
  \includegraphics[width=4in]{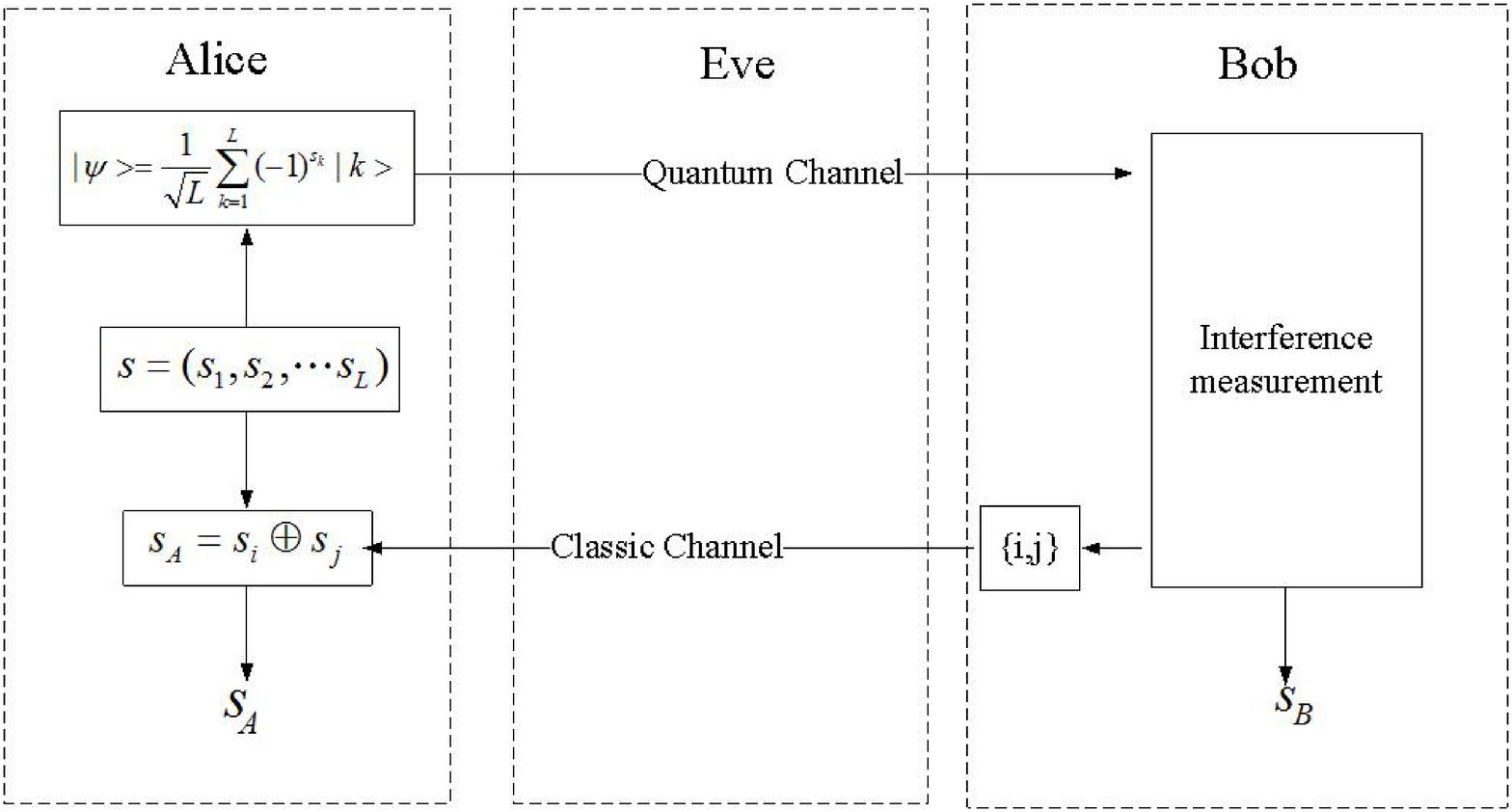}
  \caption{\textbf{Diagram of RRDPS protocol}}
  \label{1}
\end{figure}

The secure key length of QKD protocol is given by \cite{4}
\begin{equation}
G = N[1 - f \cdot h({e_{bit}}) - h({e_{phase}})]
\label{equ1}
\end{equation}
where $N$ is the sifted key length, $f$ corresponds to the efficiency of error correction, and  $h(e) =-e\log e-(1-e)\log (1-e)$ is the binary Shannon entropy function. Moreover, ${e_{bit}}$ and ${e_{phase}}$ are the bit error rate and phase error rate respectively.

As for RRDPS protocol, the phase error rate depends on the preparation of quantum states rather than the transmission process. When the number of photons in a packet is no more than an integer ${v_{th}}({v_{th}} < \frac{{L-1}}{2})$, in the analysis of Sasaki et al., the phase error rate can be bounded by ${{{v_{th}}} \mathord{\left/
 {\vphantom {{{v_{th}}} {(L - 1}}} \right.
 \kern-\nulldelimiterspace} {(L - 1}})$ \cite{11}. While they assume that Eve completely knows the sifted key bits from the rest part($v>{v_{th}}$).

Zhang et al.\cite{12} improved the phase error estimation by considering the encoding details of the quantum signal. The phase error rate is given by
\begin{gather}
e_{ph}^n = \frac{{1 - {{(1 - 2/L)}^n}}}{2}.
\end{gather}
 When the number of photons satisfies $v < {v_{th}}$, phase error rate can be bounded by $\frac{{1 - {{(1 - {2 \mathord{\left/{\vphantom {2 L}} \right. \kern-\nulldelimiterspace} L})}^{{{\rm{v}}_{th}}}}}}{2}$, which accords a tighter and more reasonable bound on the phase error rate.

Hence, the improved phase error rate estimation of source satisfying   $\Pr (v > {v_{th}}) \le {e_{src}}$ is expressed by\cite{11, 12}
\begin{gather}
 {{\rm{e}}_{ph}} = \frac{{{e_{src}}}}{Q} + (1 - \frac{{{e_{src}}}}{Q})\frac{{1 - {{(1 - {2 \mathord{\left/
 {\vphantom {2 L}} \right.
 \kern-\nulldelimiterspace} L})}^{{{\rm{v}}_{th}}}}}}{2}
\end{gather}
 where Q is the empirical rate of detection  $Q = {N \mathord{\left/{\vphantom {N {{N_{em}}}}} \right. \kern-\nulldelimiterspace} {{N_{em}}}}$, and ${N_{em}}$ corresponds to the number of packets emitted from Alice. There are  ${N_{em}}{e_{src}}$ rounds satisfying the number of photons  $v>{v_{th}}$, which are regarded as a phase error in the worst case scenario. So the first term of this equation stands for the fraction of phase error rate where the photons in a packet have exceeded ${v_{th}}$, and the second one refers to that of packets whose photons are no larger than ${v_{th}}$.

 On account of this classification, the secure key rate per packet can be calculated to be\cite{11}
\begin{gather}
R = (Q-{e_{src}})(1 - f \cdot h({e_{bit}}) - h(\frac{{1 - {{(1 - {2 \mathord{\left/ {\vphantom {2 L}} \right. \kern-\nulldelimiterspace} L})}^{{v_{th}}}}}}{2})) -{e_{src}} \cdot f \cdot h({e_{bit}}).
\label{equ7}
 \end{gather}

 The former part is the contribution of packets containing photons exceeding ${v_{th}}$, while the latter one is that of packets containing photons no more than  ${v_{th}}$. From this equation, it is clear that the larger L is, the higher secure key rate of per packet we can obtain.

\textbf{Finite decoy-state RRDPS protocol}

Decoy-state method\cite{17} has been proposed as a useful method to improve the performance of QKD protocols when using an imperfect single-photon source. Similarly, the decoy-state for RRDPS protocol also have been employed\cite{12}. Here, we review the idea of decoy-state RRDPS.

Denote ${Q_{L\mu }} = \sum\limits_{n = 0}^\infty  {{Y_n}{P_{L\mu }}(n)}$ to be the overall gain when the source intensity is $L\mu $, and ${H_{PA}}$ to be the ratio of key rate that is sacrificed in privacy amplification. According to Eq.(\ref{equ1}), the final key rate of per packet can also be written as \cite{12}
\begin{equation}
R={Q_{L\mu}}[1 - f \cdot h({e_{bit}})-{H_{PA}}].
\label{equ2}
\end{equation}

 Define ${Y_n}$  to be the yield of n-photon state, which means the conditional probability of a detection event at Bob side when Alice sends out n-photon state. Note that ${Y_0}$ is the background rate, including the detector dark count and other background contributions. Denote ${P_\mu }(n)$  to be the possibility of $n$ photons when the mean number is $\mu$. The amount of key loss can be calculated as\cite{12}
\begin{equation}
{Q_{L\mu }}{H_{PA}} = \sum\limits_{{\rm{n}} = 0}^\infty  {{Y_n}} {P_{L\mu }}(n)h({e_{ph}^n}).
\label{equ8}
\end{equation}

  ${Y_n}$ cannot be measured directly, but we can accurately estimate it with infinite decoy states. Without the interference of Eve, it is given
  by\cite{20}
 \begin{equation}
 {Y_n} = 1 - (1 - {Y_0}){(1 - \eta )^n}.
 \label{equ9}
 \end{equation}

  And the error rate ${e_n}$  can be obtained from\cite{20}
 \begin{equation}{{\rm{e}}_n} = \frac{{{e_0}{Y_0} + {e_d}(1 - {Y_0})[1 - {{(1 - \eta )}^n}]}}{{{Y_n}}}.
 \label{equ10}
 \end{equation}

  In practice, the number of decoy states cannot be chosen freely. As the number of intensity increases, the enforcement of the protocol may be more challengeable. So it is worthy to study finite decoy-state method. For practical implementations, since contributions from states with large photon numbers are negligible comparing with those from small photon numbers, only a few decoy states will be sufficient. The final key rate can be rewritten as
\begin{equation}
 R = \sum\limits_{n = 0}^{n_{th}}  {{Y_n}{P_{L\mu }}} (n)(1 - f \cdot h(e_{bit}^n) - h(e_{ph}^n))
\end{equation}
where ${Y_n}$  is defined to be the yield of an n-photon state, ${P_{L\mu }}(n)$ refers to the possibility of n photons when the mean number is $L\mu$, $f$ corresponds to the efficiency of error correction, and $h(e)=-e\log e-(1-e)\log(1-e)$ is the binary Shannon entropy function. $n_{th}$ is the threshold of the photon numbers that are  efficient. Define $e_{bit}^n$, $e_{ph}^n$ to be the bit error rate and phase error rate of n photons respectively. ${Y_n}$ and $e_{bit}^n$ need to be estimated by using the decoy states method.

In BB84, a protocol with two decoy states, the vacuum and a weak decoy state, only estimating the yield and error rate of single photon,
asymptotically approaches the theoretical limit of infinite decoy-state protocol\cite{20}. Since multi-photons are also contributed to the
secret key in RRDPS, we propose finite decoy-state to estimate ${Y_1}, {Y_2}, {Y_3}$ and ${e_1}, {e_2}, {e_3}$, that is $n_{th}=3$. Here we set an
example of WCPs to show the calculation of these  parameters.

\textbf{Numerical simulation}

To describe a practical system, a widely used fiber-based setup model is needed. When the laser source is modeled WCPs, the density matrix of the state can be given by\cite{21}
\begin{gather}
{\rho_\mu}=\int_0^{2\pi}{\frac{{d\theta }}{{2\pi }}} |\sqrt \mu  {e^{j\varphi }} >  < \sqrt \mu  {e^{j\varphi }}| = \sum\limits_{n = 0}^\infty  {{e^{ - \mu }}\frac{{{\mu ^n}}}{{n!}}} |n >  < n|
\label{equ3}
\end{gather}
where $|0><0|$ is the vacuum state and $|n><n|$ is the density matrix of the $n$-photon state.

HSPS, like the commonly used WCPs, is also within reach of current technology as another candidate of the perfect single-photon source.
Given a two-mode state of the form\cite{21}
\begin{gather}
{(\cosh \chi )^{ - 1}}\sum\limits_{n = 0}^\infty  {(\tanh } \chi {)^n}{e^{in\theta }}|n, n >
\label{equ4}
\end{gather}

Set the intensity of the source  $\mu$ to ${\sinh ^2}\chi$, then the above description simplifies to\cite{22}
\begin{gather}
\sum\limits_{n = 0}^\infty  {\sqrt {\frac{{{\mu ^n}}}{{{{(1 + \mu )}^{n + 1}}}}} } {e^{in\theta }}|n, n >.
\label{equ5}
\end{gather}

After triggering out one of a photon pair, the other mode is basically a field of distribution\cite{19}
\begin{gather}
{p_x} = \frac{1}{{{P_{post}}(\mu)}}\{ \frac{{{d_A}}}{{1 + \mu}}|0 >  < 0| + \sum\limits_{n = 1}^\infty  {[1 - {{(1 - {\eta _A})}^n}]}  \cdot \frac{{{\mu^n}}}{{{{(1 + \mu)}^{n + 1}}}}|n >  < n|\}
\label{equ6}
\end{gather}
where $\mu$  is the mean photon number of one mode, ${\eta _A}$ ,  ${d_A}$account for the detection efficiency and dark count rate of detector respectively. The post-selection probability is${P_{post}}(x) = \frac{{{d_A}}}{{1 + x}} + \frac{{x{\eta _A}}}{{1 + x{\eta _A}}}$.

Considering the distribution of photons in Eq.\ref{equ3}, the gain and QBER for using WCPs can be calculated by\cite{20}
\begin{gather}
{Q_{L\mu }} = \sum {{Y_n}\frac{{{{(L\mu )}^n}}}{{n!}}{e^{ - L\mu }}} = {Y_0} + (1 - {Y_0})(1 - {e^{ - L\eta \mu }})\\
{E_{L\mu }}{Q_{L\mu }} = \sum {{e_n}{Y_n}\frac{{{{(L\mu )}^{\rm{n}}}}}{{{{(1 + L\mu )}^{{\rm{n + 1}}}}}}} = {e_0}{Y_0} + {e_d}(1 - {Y_0})(1 - {e^{ - L\eta \mu }})
\end{gather}

Based on the formula above, we use the following for the error rate\cite{12}
\begin{gather}
{e_{bit}} = \frac{{{E_{L\mu }}{Q_{L\mu }}}}{{{Q_{L\mu }}}} = \frac{{{e_0}{Y_0} + {e_d}(1 - {Y_0})(1 - {e^{ - L\eta \mu }})}}{{{Y_0} + (1 - {Y_0})(1 - {e^{ - L\eta \mu }})}}
\end{gather}

Denote the loss coefficient in the quantum channel to be $\alpha$. For an optical-fiber-based system, the relationship between the transmission distance $d$ and the overall transmittance $\eta$ is\cite{20}
 \begin{gather}
 d=-\frac{{10{{\log}_{10}}\eta }}{\alpha}.
\end{gather}

From these formulas, the overall gain and bit error rate for HSPS can be given by
\begin{gather}
\nonumber{Q_{L\mu }} = \sum {{Y_n}{P_{L\mu }}(n)}
\\\nonumber  = {Y_0} \cdot \frac{{{d_A}}}{{(1 + L\mu ) \cdot {P_{post}}(L\mu )}} + \sum\limits_{n = 1}^\infty  {[1 - (1 - {Y_0}){{(1 - \eta )}^n}] \cdot [1 - {{(1 - {\eta _A})}^n}]}\cdot
 \frac{{{{(L\mu )}^n}}}{{{{(1 + L\mu )}^{n + 1}} \cdot {P_{post}}(L\mu )}}
\\ = \frac{{{d_A}{Y_0}(1 + L\mu {\eta _A}) + L\mu {\eta _A}(1 + L\mu )}}{{{d_A}(1 + L\mu {\eta _A}) + L\mu {\eta _A}(1 + L\mu )}}
- \frac{{L\mu (1 - {Y_0})(1 + L\mu {\eta _A})(1 - \eta )}}{{[{d_A}(1 + L\mu {\eta _A}) + L\mu {\eta _A}(1 + L\mu )](1 + L\mu \eta )}}
\\\nonumber+ \frac{{L\mu (1 - {Y_0})(1 + L\mu {\eta _A})(1 - \eta )(1 - {\eta _A})}}{{[{d_A}(1 + L\mu {\eta _A})+ L\mu {\eta _A}(1 + L\mu )](1 + L\mu \eta  + L\mu {\eta _A} - L\mu \eta {\eta _A})}}
\end{gather}
\begin{gather}
\nonumber{E_{L\mu }}{Q_{L\mu }} = \sum {{e_n}{Y_n}} {P_\mu }(n)
 \\\nonumber= {e_0}{Y_0} \cdot \frac{{{d_A}}}{{(1 + L\mu ) \cdot {P_{post}}(L\mu )}} + \sum\limits_{n = 1}^\infty  {\{ [1 - {{(1 - \eta )}^n}]\}  \cdot [1 - {{(1 - {\eta _A})}^n}]}\cdot \frac{{{{(L\mu )}^n}}}{{{{(1 + L\mu )}^{n + 1}} \cdot {P_{post}}(L\mu )}}
   \\ = \frac{{{d_A}{e_0}{Y_0}(1 + L\mu {\eta _A}) + L\mu {\eta _A}(1 + L\mu )[{e_0}{Y_0} + {e_d}(1 - {Y_0})]}}{{{d_A}(1 + L\mu {\eta _A}) + L\mu {\eta _A}(1 + L\mu )}}
   \\\nonumber -\frac{{{e_d}L\mu (1 - {Y_0})(1 + L\mu {\eta _A})(1 - \eta )}}{{({d_A}(1 + L\mu {\eta _A}) + L\mu {\eta _A}(1 + L\mu ))(1 + L\mu \eta )}}
\\\nonumber+ \frac{{{e_d}L\mu (1 - {Y_0})(1 + L\mu {\eta _A})(1 - \eta )(1 - {\eta _A})}}{{({d_A}(1 + L\mu {\eta _A}) + L\mu {\eta _A}(1 + L\mu ))(1 + L\mu \eta  + L\mu {\eta _A} - L\mu \eta {\eta _A})}}
\end{gather}

Key rate per packet, $R$, represents the average net production length of secure key in a fixed length packet. The final asymptotic secret key rate per packet can be given by equation (5). For a fixed length packet, there are optimized mean photon numbers, $\mu $, and thresholds of photons in each packet, ${v_{th}}$, in the process of transmission. To clarify the choice of $\mu $ and ${v_{th}}$ intuitionally, we show the relationship among $R$, $\mu $  and ${v_{th}}$ as following Figure 2. The length of each packet is fixed at 128. The channel transmission is $10^{-5}$. By changing values of two variables, we can observe the affect of variables on optimization results of final key rate.

 \begin{figure}[ht]
  \centering
  \includegraphics[width=4in]{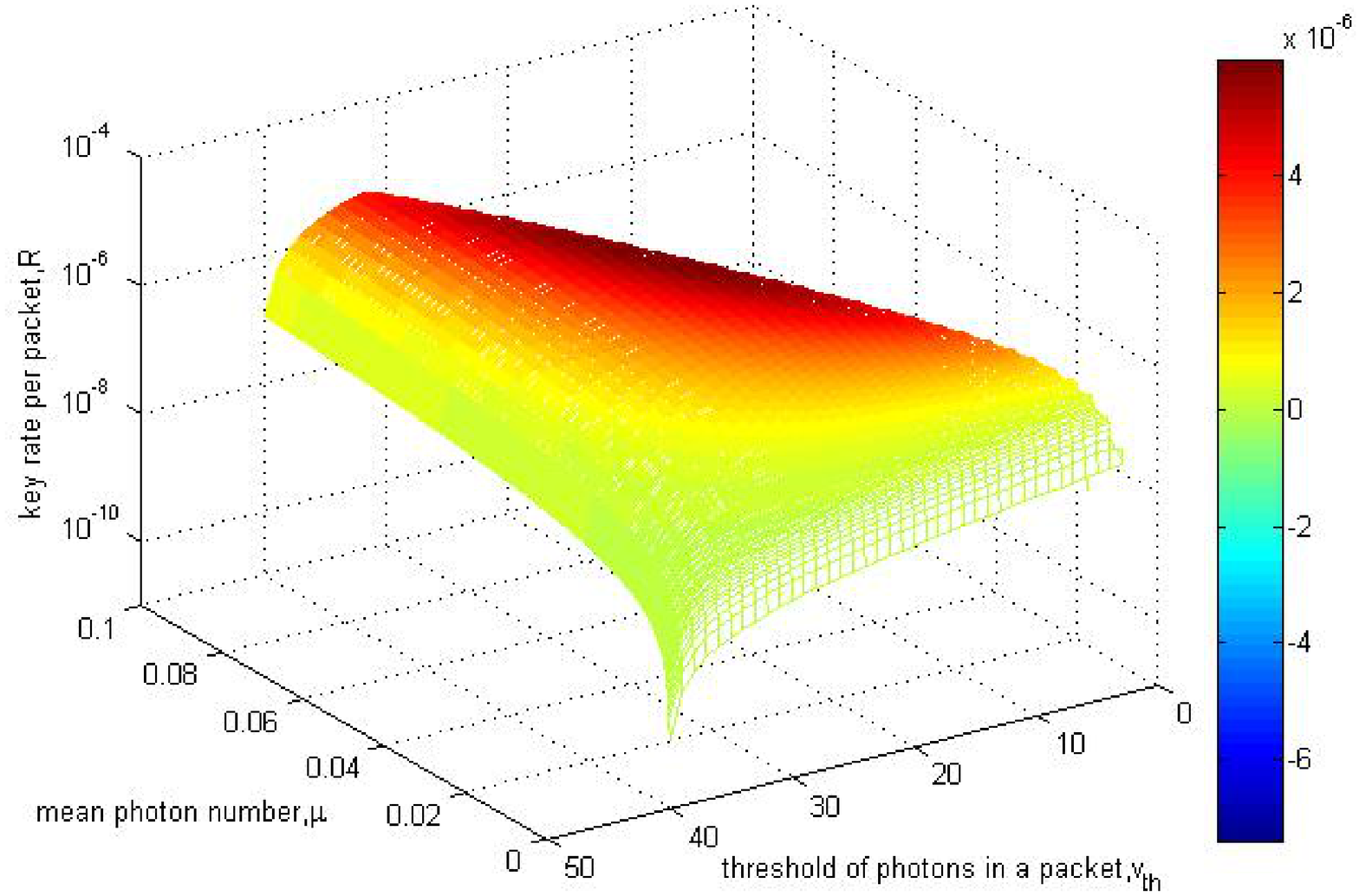}
  \caption{\textbf{(Color online) Key rates versus mean photon number and threshold of photons in a packet.} }
  \label{2}
\end{figure}

Various colors indicate diverse values of key rate per packet. By looking at the graph, all the major values will show up clearly. It is obvious that key rate per packet has a maximum point, where corresponding mean photon number and photons in each packet can be regarded as optimal ones. So in the following simulation we adopt these optimal values for each condition to obtain better key rates.

\begin{table}
\begin{center}
 \caption{\label{table1}Experimental parameters for simulation}
\begin{tabular}{c c c c c}
 \hline
 ${\eta _{\rm{A}}}$ &${d_{\rm{A}}}$&${e_d}$&$\alpha $ & $f$\\
 \hline
 0.045	&$1.7 \times {10^{ - 6}}$ &0.033&	0.2dB/km &1.16\\
  \hline
 \end{tabular}
 \end{center}
 \end{table}

Let $Q$ be the empirical rate of detection about  channel transmission $\eta $. For the use of WPCs and HSPS, the key rates per packet with different packet length and bit error rate, as a function of channel transmission, is illustrated in Figure 3. Parameters used are listed in Table 1.

\begin{figure}
 \begin{center}
  \includegraphics[width=6in]{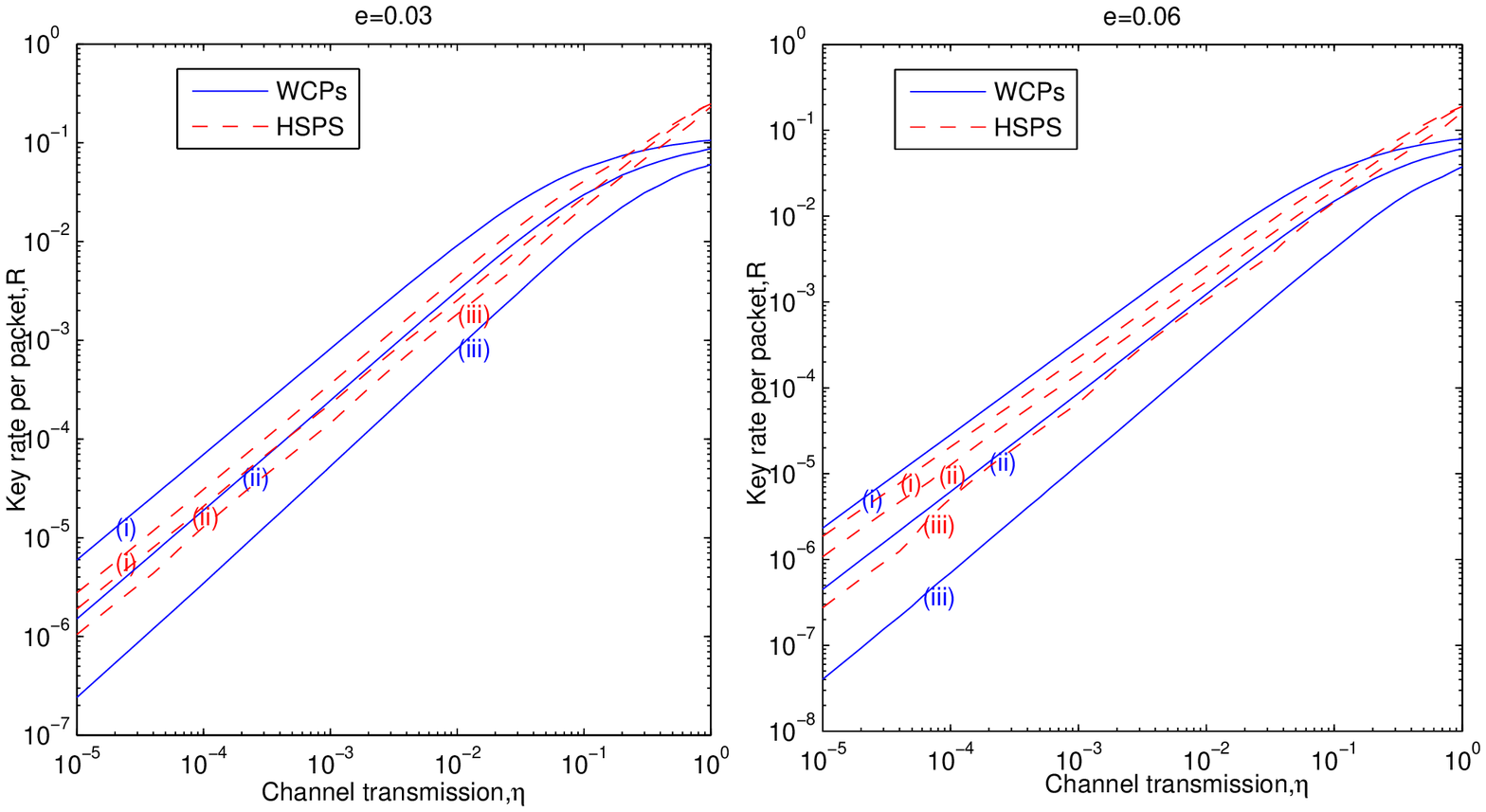}
  \caption{\textbf{(Color online) Comparison for RRDPS protocol using WCPs and HSPS. }The solid curves represent the key rates per packet for using WCPs, the dashed curves stand for key rates for using HSPS. Lines labeled (i)-(iii) characterize the protocol with L=128, 64, 32. The  error rate is 0.03 and 0.06, respectively. The choices of ${v_{th}}$  and the mean photon number $\mu $  are optimized.}
  \label{3}
  \end{center}
\end{figure}

From the results shown in Figure 3, key rates per packet increase as the variation of channel transmission. By simulation, we can see that the key rate or WCPs is mostly larger than that for HSPS when $L=128$. Two sources perform similarly if $L$ is equal to 64 and $e=0.03$. When $L$ equals to 32, HSPS performers better. In a word, WCPs is more suitable for larger packet length while HSPS performs better with smaller packet.

According to the distribution of each source, we can obtain the probability of emitting different numbers of photons among WCPs and HSPS. For a better interpretation of this fact, we take a simple comparison in Figure 4.

\begin{figure}[ht]
\centering
  \includegraphics[width=7in]{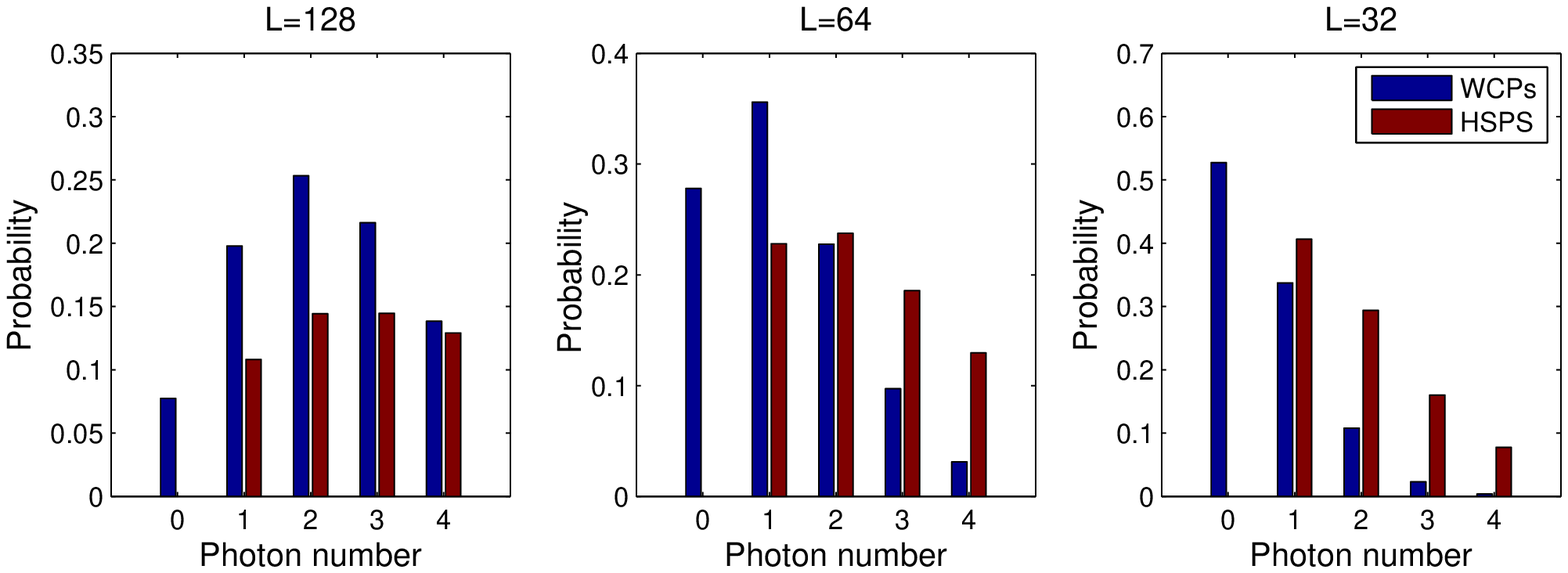}
  \caption{\textbf{(Color online) A comparison for the probability of emitting different numbers of photons between WCPs and HSPS. }Here parameters for HSPS are listed in Table~\ref{table1}, while the intensity is 0.02 for both WCPs and HSPS.}
  \label{0}
\end{figure}

It is apparent that the results of Figure 4 and Figure 3 are coincident as a whole. Roughly speaking, when $L=128$, the summation of photons contributing to secure key rate in WCPs is larger than that in HSPS. These two summations are equal to each other while $L=64$, then the former one is less than the latter one when $L=32$.

Applying the expressions of $Q_{L\mu}$ and $e_{bit}$ to Eq.\ref{equ7}, we can numerically compare the performance of the two sources with decoy-state and without decoy-state. In order to give a faithful estimation, we employ a reasonable model to forecast the result, taking example by a typical QKD system\cite{7}. Here, we fix $L$ at 32, and other parameters used are listed in Table 1.

Combing Eq.\ref{equ2}, Eq.\ref{equ8}, \ref{equ9}, \ref{equ10}, we can calculate the final key generation rate  of RRDPS protocol with and without decoy-state for WCPs and HSPS. For convenience of comparing, we use the same parameters as above in Table 1. Our simulation results are shown in Figure 5.

\begin{figure}
  \centering
  \includegraphics[width=4in]{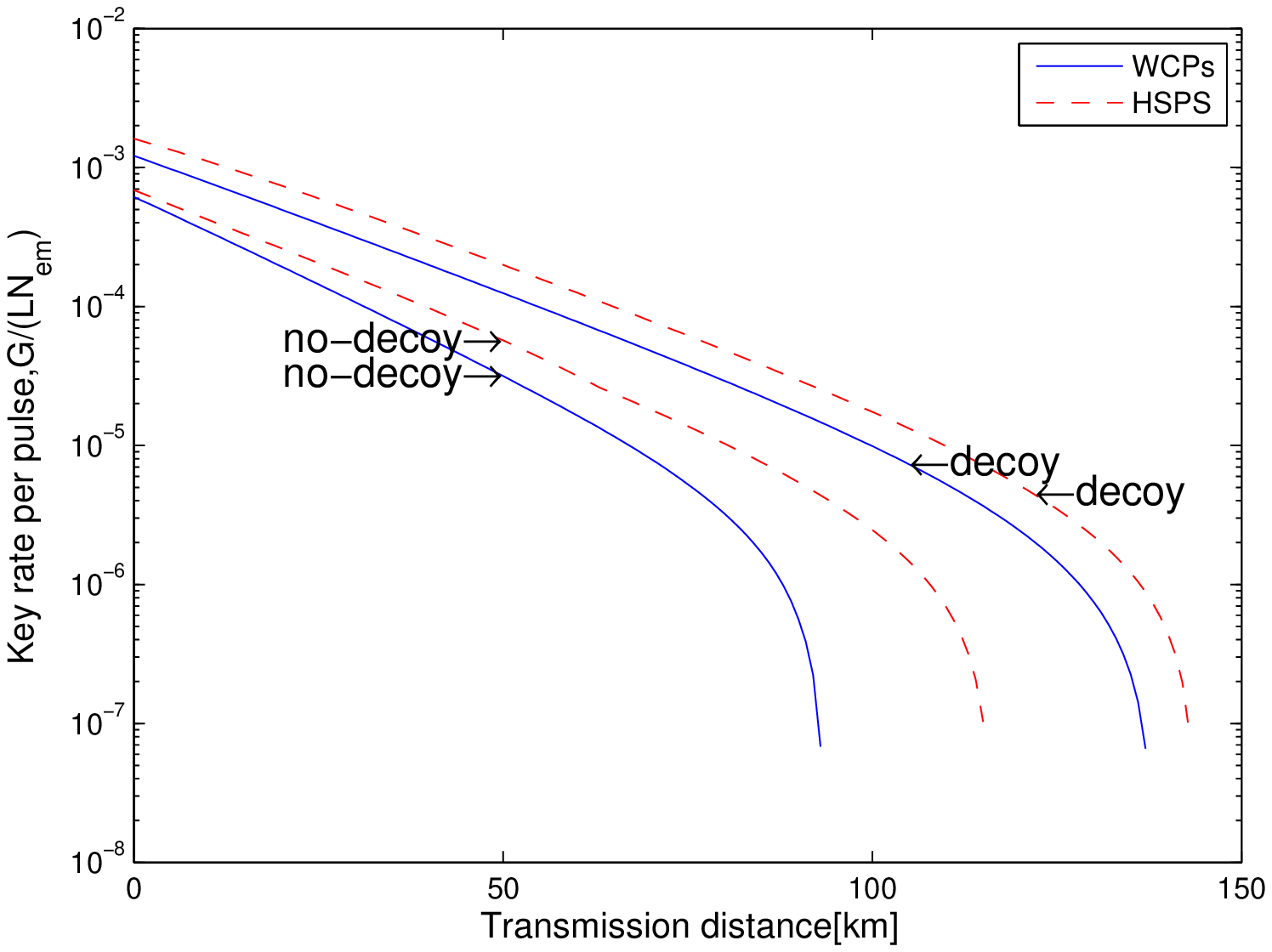}
  \caption{\textbf{(Color online) Final key rate without and with decoy states. }The solid  curves represent the key rates per packet for using WCPs, and the dashed curves stand for key rates for using HSPS. The length of each packet is fixed at 32. The left two curves are the key rates of RRDPS protocol with no decoy states, and the other two are those of the protocol with decoy states.}
  \label{4}
\end{figure}

The decoy-state method is often used to improve the secure key rate and transmission distance in conventional protocol. The simulation result in Figure 5 clearly demonstrates that this method is also effective for the RRPDS protocol. The performance of HSPS is much better than WCPs under these two circumstances.

For the case of WCPs, we compare decoy-state methods for RRDPS protocol with different intensities: infinite decoy states, two-decoy-state method in which we just estimate ${Y_1}$  and $e_1$, three-intensity decoy-state estimating  ${Y_1}$, ${Y_2}$, $e_1$ and $e_2$, four-decoy-states estimating ${Y_1}$, ${Y_2}$, ${Y_3}$, $e_1$, $e_2$ and $e_3$.(see the Methods) Then, the comparison between the estimated values and asymptotic values of yields and error rates are shown in Figure 6.

\begin{figure}[ht]
  \centering
  \includegraphics[width=6in]{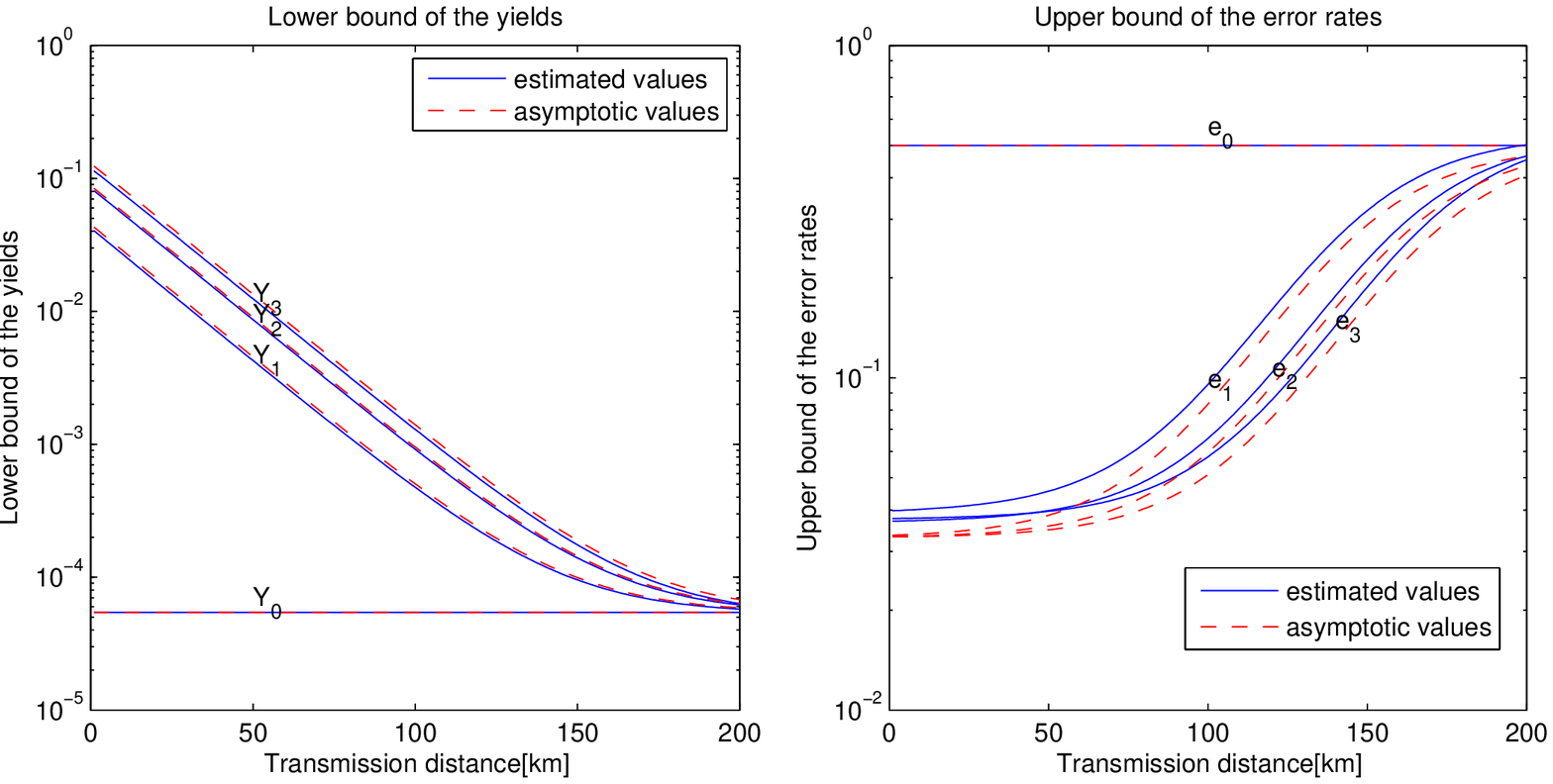}
  \caption{ \textbf{(Color online) The estimated bounds and asymptotic bounds of yields and error rates.}The solid  lines represent estimated bounds of yields and error rates, and the dashed lines stand for asymptotic bounds of yields and error rates.}
  \label{5}
\end{figure}

From the graph it is evident that the estimated values of yields are infinite approaching the asymptotic values. The estimated error rates are very close to the asymptotic values. Since we have presented partial lower bounds of yield and upper bounds of QBER, we will give an example to show that, even in the case of finite, the performance of our method is close to that of the infinite decoy method. We use the key parameter listed in Table I. Here, we also fix $L$ at 32, and optimize $\mu$ to obtain the maximum transmission distance. The simulation result indicates that three decoy states are sufficient for the RRDPS protocol, which is shown in Figure 7.

\begin{figure}[ht]
  \centering
  \includegraphics[width=4in]{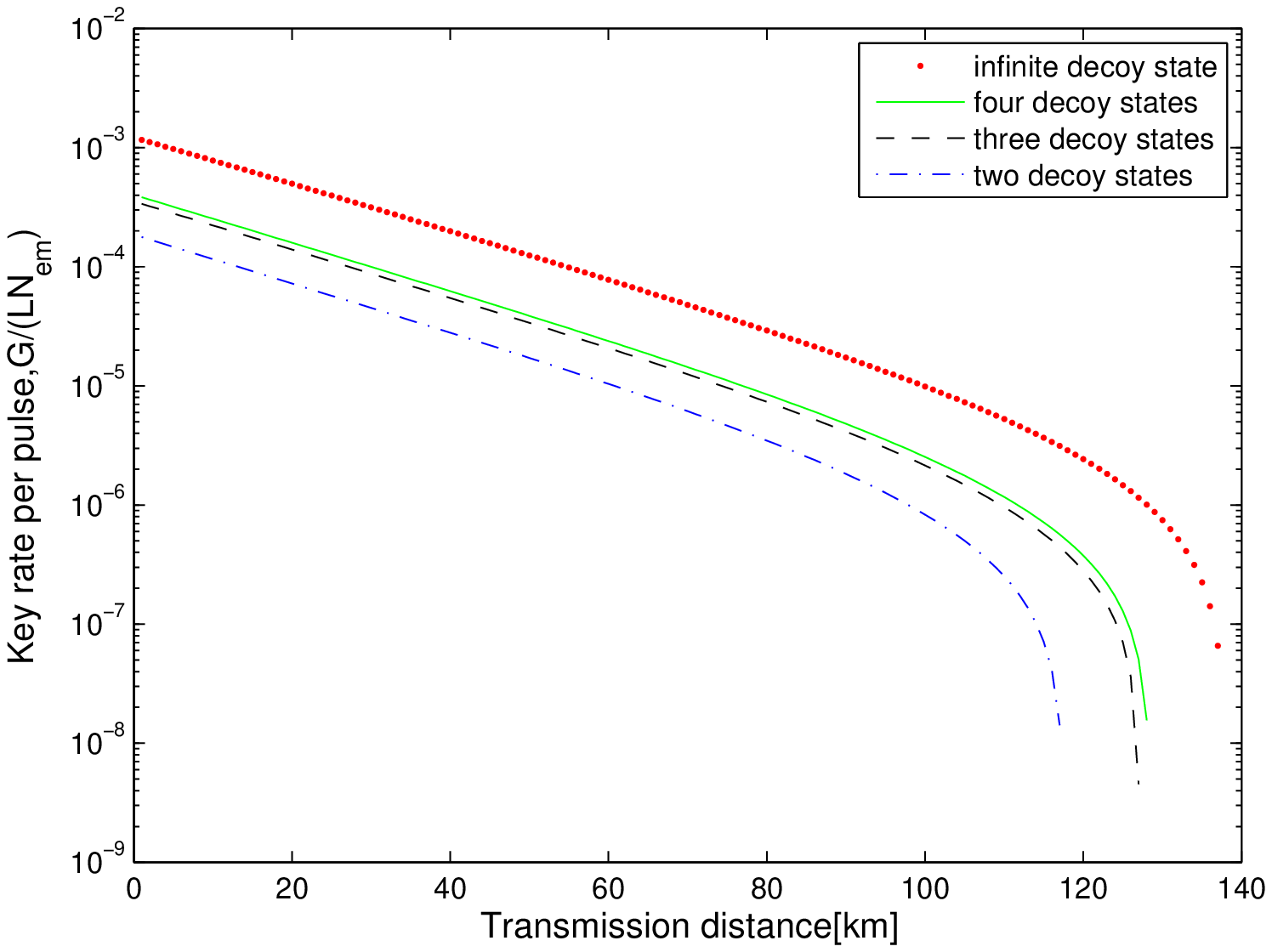}
  \caption{ \textbf{(Color online) Final key rate for using infinite and finite decoy states. }The dotted curve accounts for the key rate per packet for infinite decoy-state, and the solid curve stands for the four-intensity decoy-state method to estimate${Y_1}$, ${Y_2}$, ${Y_3}$, $e_1$, $e_2$ and $e_3$. The dashed curve represents for three-intensity decoy-state method estimating  ${Y_1}$, ${Y_2}$, $e_1$  and$e_2$, and the dash-dotted one is on behalf of the two-intensity decoy-state method just estimating ${Y_1}$  and$e_1$. The choices of the mean photon number $\mu$  are optimized.}
  \label{6}
\end{figure}

In the decoy-state method of BB84 protocol, only  $Y_1$ and  $e_1$ are needed to estimate, so a few decoy states will be sufficient. This is because contributions from states with large photon numbers are negligible comparing with those from small photon numbers. From the results, we can see that contributions from two-photon state are considerable, while those from three photons are not obvious. Therefore, it is dispensable to estimate yield and QBER when the photon number is more than three.

\textbf{Discussions}

From the comparison between WCPs and HSPS, it is clear that they perform differently with various packet lengths. WCPs is more suitable for larger packet length while HSPS performs better with smaller packet. We anticipate that this result can guide the use of conventional lasers. In current practice, attenuated lasers emitting WCPs are employed in most quantum system. The technology on how to efficiently obtain HSPS has been developed to a high level. It is pragmatic to experimentally test the conclusions from numerical simulations.  For practical implementations, we put forward finite decoy-state protocol. Since the key rate of the finite decoy-state method is close to the infinite one, our protocol can be regarded as a choice for the practical experiment of RRDPS. Taking cost and and technological feasibility into account, our scheme can be promising implementation of quantum cryptography. Compared with existing QKD protocol, there are still many aspects to improve for RRDPS. For example, the finite length of key\cite{23} and its statistical fluctuations can significantly affect the security of RRDPS protocol, which is of special concern.

\section*{Methods}
\textbf{Lower bound of ${Y_1}, {Y_2}, {Y_3}$\\}

Suppose Alice and Bob choose a signal state with intensity  $L\mu$ and four decoy states whose intensities are $L{v_1}, L{v_2}, L{v_3}, L{v_4}$, which satisfies
\begin{gather}
L{v_1} \ge L{v_2} \ge L{v_3} \ge L{v_4} \ge 0, L\mu  > L{v_1} + L{v_2} + L{v_3} + L{v_4}.
\end{gather}

One can obtain the following gains and quantum bit error rates for the signal and decoy states
\begin{gather}
\nonumber{Q_{L\mu }}{{\rm{e}}^{L\mu }} = \sum\limits_{n = 0}^\infty  {{Y_n}\frac{{{{(L\mu )}^n}}}{{n!}}} \qquad
{E_{L\mu }}{Q_{L\mu }}{e^{L\mu }} = \sum\limits_{n = 0}^\infty  {{e_n}{Y_n}\frac{{{{(L\mu )}^n}}}{{n!}}} \\ \nonumber
{Q_{L{v_1}}}{e^{L{v_1}}} = \sum\limits_{n = 0}^\infty  {{Y_n}\frac{{{{(L{v_1})}^n}}}{{n!}}} \qquad
{E_{Lv}}_{_1}{Q_{Lv}}_{_1}{e^{L{v_1}}} = \sum\limits_{n = 0}^\infty  {{e_n}{Y_n}\frac{{{{(L{v_1})}^n}}}{{n!}}} \\
{Q_{L{v_2}}}{e^{L{v_2}}} = \sum\limits_{n = 0}^\infty  {{Y_n}\frac{{{{(L{v_2})}^n}}}{{n!}}} \qquad  {E_{Lv}}_{_2}{Q_{Lv}}_{_2}{e^{L{v_2}}} = \sum\limits_{n = 0}^\infty  {{e_n}{Y_n}\frac{{{{(L{v_2})}^n}}}{{n!}}} \\\nonumber
{Q_{L{v_3}}}{e^{L{v_3}}} = \sum\limits_{n = 0}^\infty  {{Y_n}\frac{{{{(L{v_3})}^n}}}{{n!}}} \qquad {E_{Lv}}_{_3}{Q_{Lv}}_{_3}{e^{L{v_3}}} = \sum\limits_{n = 0}^\infty  {{e_n}{Y_n}\frac{{{{(L{v_3})}^n}}}{{n!}}}\\\nonumber
{Q_{L{v_4}}}{e^{L{v_4}}} = \sum\limits_{n = 0}^\infty  {{Y_n}\frac{{{{(L{v_4})}^n}}}{{n!}}} \qquad {E_{Lv}}_{_4}{Q_{Lv}}_{_4}{e^{L{v_4}}} = \sum\limits_{n = 0}^\infty  {{e_n}{Y_n}\frac{{{{(L{v_4})}^n}}}{{n!}}}
\end{gather}

Given such equations, how can we obtain a tight lower bound on $R$. This is the main problem for decoy-state protocols.

Based on  ${v_1}{Q_{L{v_2}}}{e^{L{v_2}}} - {v_2}{Q_{L{v_1}}}{e^{L{v_1}}}$, a crude lower bound of ${Y_0}$  can be chosen by\cite{20}
\begin{gather}
{Y_0} \ge {Y_0}^L = \max (\frac{{{v_1}{Q_{L{v_2}}}{e^{L{v_2}}} - {v_2}{Q_{L{v_1}}}{e^{L{v_1}}}}}{{{v_1} - {v_2}}}, 0)
\end{gather}
when  $L{v_2} = 0$ the equality sign will hold.

By   ${Q_{L{v_1}}}{e^{L{v_1}}} - {Q_{L{v_2}}}{e^{L{v_2}}}$, it is clearly that
\begin{gather}
{Q_{L{v_1}}}{e^{L{v_1}}} - {Q_{L{v_2}}}{e^{L{v_2}}} \le {Y_1}(L{v_1}
 - L{v_2}) + \frac{{{{(L{v_1})}^2} - {{(L{v_2})}^2}}}{{{{(L\mu )}^2}}}({Q_{L\mu }}{e^{L\mu }} - Y_0^L - {Y_1}L\mu).
 \end{gather}

Consequently, we obtain a minimum value of  ${Y_1}$\cite{20}
\begin{gather}
{Y_1} \ge {Y_1}^L = \frac{\mu }{{L(\mu {v_1} - \mu {v_2} - {v_1}^2 + {v_2}^2)}}({Q_{L{v_1}}}{e^{L{v_1}}} - {Q_{L{v_2}}}{e^{L{v_2}}} - \frac{{{v_1}^2 - {v_2}^2}}{{{\mu ^2}}}({Q_{L\mu }}{e^{L\mu }} - Y_0^L)).
\end{gather}
where $Y_0^L$  is the lower bound of  ${Y_0}$ calculated earlier.

Combine these two equations
\begin{eqnarray}
{Q_{L{v_1}}}{e^{L{v_1}}} - {Q_{L{v_2}}}{e^{L{v_2}}} = {Y_1}(L{v_1} - L{v_2}) + \frac{{{Y_2}}}{{2!}}({(L{v_1})^2} - {(L{v_2})^2}) + \frac{{{Y_3}}}{{3!}}({(L{v_1})^3} - {(L{v_2})^3}) +  \cdots \\
{Q_{L{v_2}}}{e^{L{v_2}}} - {Q_{L{v_3}}}{e^{L{v_3}}} = {Y_1}(L{v_2} - L{v_3}) + \frac{{{Y_2}}}{{2!}}({(L{v_2})^2} - {(L{v_3})^2}) + \frac{{{Y_3}}}{{3!}}({(L{v_2})^3} - {(L{v_3})^3}) +  \cdots
\end{eqnarray}

With $({Q_{L{v_1}}}{e^{L{v_1}}} - {Q_{L{v_2}}}{e^{L{v_2}}}) \times ({v_2} - {v_3}) -  ({Q_{L{v_2}}}{e^{L{v_2}}} - {Q_{L{v_3}}}{e^{L{v_3}}}) \times ({v_1} - {v_2})$, we also have
\begin{gather}
 \nonumber ({v_2} - {v_3}){Q_{{v_1}}}{e^{{v_1}}} -({v_1}-{v_3}){Q_{{v_2}}}{e^{{v_2}}} + ({v_1} - {v_2}){Q_{{v_3}}}{e^{{v_3}}} \le \frac{{{Y_2}}}{2}{L^2}({v_2} - {v_3})({v_1} - {v_3})({v_1} - {v_2})\\
+\frac{{({v_2} - {v_3})({v_1} - {v_3})({v_1} - {v_2})({v_1} + {v_2} + {v_3})} }{{{{\mu}^3}}}({Q_{L\mu }}{e^{L\mu }} - Y_0^L - {Y_1}^L{L\mu}- \frac{{{Y_2}{{(L\mu )}^2}}}{2})
\end{gather}

Therefore, we can bound ${Y_2}$  by

\begin{gather}
   \nonumber {{Y}_{2}}\ge {{Y}_{2}}^{L}=\frac{2\mu [({{v}_{2}}-{{v}_{3}}){{Q}_{L{{v}_{1}}}}{{e}^{L{{v}_{1}}}}-({{v}_{1}}-{{v}_{3}}){{Q}_{L{{v}_{2}}}}{{e}^{L{{v}_{2}}}}+({{v}_{1}}-{{v}_{2}}){{Q}_{L{{v}_{3}}}}{{e}^{L{{v}_{3}}}}]}{{{L}^{2}}(\mu -{{v}_{1}}-{{v}_{2}}-{{v}_{3}})({{v}_{2}}-{{v}_{3}})({{v}_{1}}-{{v}_{3}})({{v}_{1}}-{{v}_{2}})}
  \\-\frac{{{v}_{1}}+{{v}_{2}}+{{v}_{3}}}{{{(L\mu )}^{2}}(\mu -{{v}_{1}}-{{v}_{2}}-{{v}_{3}})}({{Q}_{L\mu }}{{e}^{L\mu }}-Y_{0}^{L}-{{Y}_{1}}^{L}L\mu)
\end{gather}

As we have known,

\begin{gather}
  \nonumber({{v}_{2}}-{{v}_{3}}){{Q}_{L{{v}_{1}}}}{{e}^{L{{v}_{1}}}}-({{v}_{1}}-{{v}_{3}}){{Q}_{L{{v}_{2}}}}{{e}^{L{{v}_{2}}}}+({{v}_{1}}-{{v}_{2}}){{Q}_{L{{v}_{3}}}}{{e}^{L{{v}_{3}}}}\\ =\frac{{{Y}_{2}}}{2!}{{L}^{2}}({{v}_{2}}-{{v}_{3}})({{v}_{1}}-{{v}_{3}})({{v}_{1}}-{{v}_{2}})+\frac{{{Y}_{3}}}{3!}{{L}^{3}}({{v}_{2}}-{{v}_{3}})({{v}_{1}}-{{v}_{2}})({{v}_{1}}-{{v}_{3}})({{v}_{1}}+{{v}_{2}}+{{v}_{3}})\\
  \nonumber+\frac{{{Y}_{4}}}{4!}{{L}^{4}}({{v}_{2}}-{{v}_{3}})({{v}_{1}}-{{v}_{2}})({{v}_{1}}-{{v}_{3}})({{v}_{1}}^{2}+{{v}_{2}}^{2}+{{v}_{3}}^{2}+{{v}_{1}}{{v}_{2}}+{{v}_{1}}{{v}_{3}}+{{v}_{2}}{{v}_{3}})+\cdots \end{gather}
\begin{gather} \nonumber({{v}_{3}}-{{v}_{4}}){{Q}_{L{{v}_{2}}}}{{e}^{L{{v}_{2}}}}-({{v}_{2}}-{{v}_{4}}){{Q}_{L{{v}_{3}}}}{{e}^{L{{v}_{3}}}}+({{v}_{2}}-{{v}_{3}}){{Q}_{L{{v}_{4}}}}{{e}^{L{{v}_{4}}}} \\
 =\frac{{{Y}_{2}}}{2!}{{L}^{2}}({{v}_{3}}-{{v}_{4}})({{v}_{2}}-{{v}_{4}})({{v}_{2}}-{{v}_{3}})+\frac{{{Y}_{3}}}{3!}{{L}^{3}}({{v}_{3}}-{{v}_{4}})({{v}_{2}}-{{v}_{3}})({{v}_{2}}-{{v}_{4}})({{v}_{2}}+{{v}_{3}}+{{v}_{4}}) \\
 \nonumber+\frac{{{Y}_{4}}}{4!}{{L}^{4}}({{v}_{3}}-{{v}_{4}})({{v}_{2}}-{{v}_{3}})({{v}_{2}}-{{v}_{4}})({{v}_{2}}^{2}+{{v}_{3}}^{2}+{{v}_{4}}^{2}+{{v}_{2}}{{v}_{3}}+{{v}_{2}}{{v}_{4}}+{{v}_{3}}{{v}_{4}})+\cdots  \end{gather}

For a similar settlement, we can also give the lower bound of  ${Y_3}$

\begin{gather}
 \nonumber {{Y}_{3}}\ge {{Y}_{3}}^{L}=\frac{3!\mu }{{{L}^{3}}(\mu -{{v}_{1}}-{{v}_{2}}-{{v}_{3}}-{{v}_{4}})}[\frac{{{Q}_{L{{v}_{1}}}}{{e}^{L{{v}_{1}}}}}{({{v}_{1}}-{{v}_{2}})({{v}_{1}}-{{v}_{3}})({{v}_{1}}-{{v}_{4}})}-\frac{{{Q}_{L{{v}_{2}}}}{{e}^{L{{v}_{2}}}}}{({{v}_{2}}-{{v}_{3}})({{v}_{1}}-{{v}_{2}})({{v}_{2}}-{{v}_{4}})} \\
+\frac{{{Q}_{L{{v}_{3}}}}{{e}^{L{{v}_{3}}}}}{({{v}_{1}}-{{v}_{3}})({{v}_{2}}-{{v}_{3}})({{v}_{3}}-{{v}_{4}})}
-\frac{{{Q}_{L{{v}_{4}}}}{{e}^{L{{v}_{4}}}}}{({{v}_{1}}-{{v}_{4}})({{v}_{2}}-{{v}_{4}})({{v}_{3}}-{{v}_{4}})}] \\\nonumber-\frac{3!({{v}_{1}}+{{v}_{2}}+{{v}_{3}}+{{v}_{4}})}{\mu -{{v}_{1}}-{{v}_{2}}-{{v}_{3}}-{{v}_{4}}}\frac{{{Q}_{L\mu }}{{e}^{L\mu }}-Y_{0}^{L}-{{Y}_{1}}^{L}L\mu -\frac{{{Y}_{2}}^{L}{{(L\mu )}^{2}}}{2!}}{{{(L\mu )}^{3}}}
\end{gather}

\textbf{Upper bound of QBER ${e_1}, {e_2}, {e_3}$}

According to the condition, the QBER of decoy state is given by\cite{20}
\begin{gather}
{E_{Lv_1}}{Q_{Lv_1}}{e^{L{v_1}}} = {e_0}{Y_0} + {e_1}L{v_1}{Y_1} + \sum\limits_{n = 2}^\infty  {{e_n}{Y_n}\frac{{{{(L{v_1})}^n}}}{{n!}}}\\
{E_{Lv_2}}{Q_{Lv_2}}{e^{L{v_2}}} = {e_0}{Y_0} + {e_1}L{v_2}{Y_1} + \sum\limits_{n = 2}^\infty  {{e_n}{Y_n}\frac{{{{(L{v_2})}^n}}}{{n!}}}
\end{gather}

The upper bound of ${e_1}$  can be obtained directly\cite{20}
\begin{equation}
{e_1} \le {e_1}^U = \frac{{{E_{Lv}}_{_1}{Q_{Lv}}_{_1}{e^{L{v_1}}} - {E_{Lv}}_{_2}{Q_{Lv}}_{_2}{e^{L{v_2}}}}}{{(L{v_1} - L{v_2}){Y_1}^L}}
\end{equation}

Combining
\begin{gather}
{E_{Lv}}_{_1}{Q_{L{v_1}}}{e^{L{v_1}}} - {E_{Lv}}_{_2}{Q_{L{v_2}}}{e^{L{v_2}}} = {e_1}{Y_1}(L{v_1} - L{v_2})
\\\nonumber+ \frac{{{e_2}{Y_2}}}{{2!}}({(L{v_1})^2} - {(L{v_2})^2}) + \frac{{{e_3}{Y_3}}}{{3!}}({(L{v_1})^3} - {(L{v_2})^3}) +  \cdots
\\
{E_{Lv}}_{_2}{Q_{L{v_2}}}{e^{L{v_2}}} - {E_{Lv}}_{_3}{Q_{L{v_3}}}{e^{L{v_3}}} = {e_1}{Y_1}(L{v_2} - L{v_3})
\\\nonumber+ \frac{{{e_2}{Y_2}}}{{2!}}({(L{v_2})^2} - {(L{v_3})^2})+ \frac{{{e_3}{Y_3}}}{{3!}}({(L{v_2})^3} - {(L{v_3})^3}) +  \cdots
\end{gather}
and solving the equality
\begin{gather}
({{v}_{2}}-{{v}_{3}}){{E}_{Lv}}_{_{1}}{{Q}_{L{{v}_{1}}}}{{e}^{L{{v}_{1}}}}-({{v}_{1}}-{{v}_{3}}){{E}_{Lv}}_{_{2}}{{Q}_{L{{v}_{2}}}}{{e}^{L{{v}_{2}}}}+({{v}_{1}}-{{v}_{2}}){{E}_{Lv}}_{_{3}}{{Q}_{L{{v}_{3}}}}{{e}^{L{{v}_{3}}}}
  \\\nonumber\ge \frac{{{e}_{2}}{{Y}_{2}}{{L}^{2}}}{2}({{v}_{2}}-{{v}_{3}})({{v}_{1}}-{{v}_{3}})({{v}_{1}}-{{v}_{2}})
\end{gather}
the upper bound of  ${e_2}$ can be further represented by
\begin{gather}
{{e}_{2}}\le {{e}_{2}}^{U}=\frac{2[({{\text{v}}_{2}}-{{v}_{3}}){{E}_{Lv}}_{_{1}}{{Q}_{L{{v}_{1}}}}{{e}^{L{{v}_{1}}}}-({{\text{v}}_{1}}-{{v}_{3}}){{E}_{Lv}}_{_{2}}{{Q}_{L{{v}_{2}}}}{{e}^{L{{v}_{2}}}}+({{v}_{1}}-{{v}_{2}}){{E}_{Lv}}_{_{3}}{{Q}_{L{{v}_{3}}}}{{e}^{L{{v}_{3}}}}]}{{{L}^{2}}{{Y}_{2}}^{L}({{\text{v}}_{2}}-{{v}_{3}})({{\text{v}}_{1}}-{{v}_{3}})({{\text{v}}_{1}}-{{v}_{2}})}
\end{gather}

For the same reason, we can obtain the equations followed,

\begin{gather}
 ({{v}_{2}}-{{v}_{3}}){{E}_{Lv}}_{_{1}}{{Q}_{L{{v}_{1}}}}{{e}^{L{{v}_{1}}}}-({{v}_{1}}-{{v}_{3}}){{E}_{Lv}}_{_{2}}{{Q}_{L{{v}_{2}}}}{{e}^{L{{v}_{2}}}}+({{v}_{1}}-{{v}_{2}}){{E}_{Lv}}_{_{3}}{{Q}_{L{{v}_{3}}}}{{e}^{L{{v}_{3}}}} \\
 \nonumber =\frac{{{e}_{2}}{{Y}_{2}}{{L}^{2}}}{2}({{v}_{2}}-{{v}_{3}})({{v}_{1}}-{{v}_{3}})({{v}_{1}}-{{v}_{2}})+\frac{{{e}_{3}}{{Y}_{3}}{{L}^{3}}}{3!}({{v}_{2}}-{{v}_{3}})({{v}_{1}}-{{v}_{3}})
 ({{v}_{1}}-{{v}_{2}})({{v}_{1}}+{{v}_{2}}+{{v}_{3}})+\cdots \\
 ({{v}_{3}}-{{v}_{4}}){{E}_{Lv}}_{_{2}}{{Q}_{L{{v}_{2}}}}{{e}^{L{{v}_{2}}}}-({{v}_{2}}-{{v}_{4}}){{E}_{Lv}}_{_{3}}{{Q}_{L{{v}_{3}}}}{{e}^{L{{v}_{3}}}}+({{v}_{2}}-{{v}_{3}}){{E}_{Lv}}_{_{4}}{{Q}_{L{{v}_{4}}}}{{e}^{L{{v}_{4}}}} \\\nonumber
  =\frac{{{e}_{2}}{{Y}_{2}}{{L}^{2}}}{2}({{v}_{3}}-{{v}_{4}})({{v}_{2}}-{{v}_{4}})({{v}_{2}}-{{v}_{3}})+\frac{{{e}_{3}}{{Y}_{3}}{{L}^{3}}}{3!}({{v}_{3}}-{{v}_{4}})({{v}_{2}}-{{v}_{4}})
  ({{v}_{2}}-{{v}_{3}})({{v}_{2}}+{{v}_{3}}+{{v}_{4}})+\cdots
\end{gather}

Then, the upper bound of $e_3$  can be shown by
\begin{gather}
 \nonumber  {{e}_{3}}\le {{e}_{3}}^{U}=\frac{3!}{{{Y}_{3}}^{L}{{L}^{3}}}[\frac{{{E}_{Lv}}_{_{1}}{{Q}_{L{{v}_{1}}}}{{e}^{L{{v}_{1}}}}}{({{v}_{1}}-{{v}_{2}})({{v}_{1}}-{{v}_{3}})({{v}_{1}}-{{v}_{4}})}-\frac{{{E}_{Lv}}_{_{2}}{{Q}_{L{{v}_{2}}}}{{e}^{L{{v}_{2}}}}}{({{v}_{2}}-{{v}_{3}})({{v}_{1}}-{{v}_{2}})({{v}_{2}}-{{v}_{4}})}
\\ +\frac{{{E}_{Lv}}_{_{3}}{{Q}_{L{{v}_{3}}}}{{e}^{L{{v}_{3}}}}}{({{v}_{1}}-{{v}_{3}})({{v}_{2}}-{{v}_{3}})({{v}_{3}}-{{v}_{4}})}
 -\frac{{{E}_{Lv}}_{_{3}}{{Q}_{L{{v}_{4}}}}{{e}^{L{{v}_{4}}}}}{({{v}_{1}}-{{v}_{4}})({{v}_{2}}-{{v}_{4}})({{v}_{3}}-{{v}_{4}})}]
\end{gather}

 \textbf{Acknowledgements}

 The authors would like to thank Zhen-Qiang Yin and Chun-Mei Zhang for their helpful discussions. This work was supported by the National Basic Research Program of China (Grant No. 2013CB338002) and the National Natural Science Foundation of China (Grants No. 11304397 and No. 61505261).

\textbf{Author Contributions}

Y. Y. Z. presented the theoretical analysis and the numerical simulations and wrote the main manuscript text. C. Z., H. W. L. and Y. W. contributed to the theoretical analysis and discussed the results. W. S. B. and M. S. J. provided essential comments on the manuscript. All authors conceived the research, and commented on the manuscript.

 \textbf{Author Information}

 The authors declare no competing financial interests.

\end{document}